\documentclass[12pt]{article}
\usepackage{mathrsfs}
\usepackage{amsfonts}
\usepackage{amsmath}
\usepackage{amssymb}
\usepackage{mathptmx}
\usepackage[dvips]{color}
\usepackage{epsfig}
\usepackage{graphicx}
\newcommand{\calA}{{\cal A}}
\newcommand{\calC}{{\cal C}}
\newcommand{\calF}{{\cal F}}
\newcommand{\calG}{{\cal G}}
\newcommand{\calH}{{\cal H}}
\newcommand{\calJ}{{\cal J}}
\newcommand{\calK}{{\cal K}}
\newcommand{\calL}{{\cal L}}
\newcommand{\calM}{{\cal M}}

\newcommand{\calW}{{\cal W}}

\newcommand{\calZ}{{\cal Z}}

\newcommand{\lam}{\lambda}

\newcommand{\Br}{\textrm{Br}}

\newcommand{\eV}{{\rm eV}}
\newcommand{\GeV}{{\rm GeV}}

\newcommand{\TeV}{{\rm TeV}}
\newcommand{\rmv}{{\rm v}}

\begin{document}
\baselineskip=16pt

\pagenumbering{arabic}

\vspace{1.0cm}

\begin{center}
{\Large\sf Flavor violating transitions of charged leptons from a
seesaw mechanism of dimension seven}
\\[10pt]
\vspace{.5 cm}

{Yi Liao$^{a,b,c}$\footnote{liaoy@nankai.edu.cn}, Guo-Zhu
Ning$^b$\footnote{ngz@mail.nankai.edu.cn}, Lu Ren$^b$}

{\it $^a$ Center for High Energy Physics, Peking University, Beijing
100871, China\\
$^b$ School of Physics, Nankai University, Tianjin 300071, China\\
$^c$ Institut f\"ur Theoretische Physik, Universit\"at Heidelberg,
Philosophenweg 16, D-69120 Heidelberg, Germany}

\vspace{2.0ex}

{\bf Abstract}

\end{center}
A mechanism has been suggested recently to generate the neutrino
mass out of a dimension-seven operator. This is expected to relieve
the tension between the occurrence of a tiny neutrino mass and the
observability of other physics effects beyond it. Such a mechanism
would inevitably entail lepton flavor violating effects. We study in
this work the radiative and purely leptonic transitions of the light
charged leptons. In so doing we make a systematic analysis of the
flavor structure by providing a convenient parametrization of the
mass matrices in terms of independent physical parameters and
diagonalizing them explicitly. We illustrate our numerical results
by sampling over two CP phases and one Yukawa coupling which are the
essential parameters in addition to the heavy lepton mass. We find
that with the stringent constraints coming from the muon decays and
the muon-electron conversion in nuclei taken into account the decays
of the tau lepton are severely suppressed in the majority of
parameter space. There exist, however, small regions in which some
tau decays can reach a level that is about 2 orders of magnitude
below their current bounds.

\begin{flushleft}


\end{flushleft}

\newpage

\section{Introduction}
\label{sec:intro}

The tiny neutrino mass and significant lepton mixing can be
incorporated in the three canonical seesaw mechanisms
\cite{type1,type2,Foot:1988aq}. From the point of view of effective
field theories they correspond to the three possible realizations at
the tree level \cite{Ma:1998dn} of the unique dimension-five
operator that induces a neutrino mass \cite{Weinberg:1979sa}. The
tininess of the neutrino mass is generally attributed to the
existence of very heavy new particles or very small couplings
between the new particles and those that we already know of. In such
a circumstance it is usually hard to detect other effects beyond the
neutrino mass.

The above tension between the occurrence of a tiny neutrino mass and
the testability of other physical phenomena can be alleviated by
postponing the appearance of higher-dimensional operators relevant
to the neutrino mass. There are two basic approaches to accomplish
this. One can compose new fields so that the operators first occur
at one \cite{Zee:1980ai}, two \cite{Zee:1985id,Babu:1988ki}, or even
three loop order \cite{Krauss:2002px}. Since the loop effects
provide additional suppressing factors besides a product of multiple
couplings, one may gain in the couplings between the new and known
particles. In the second approach, one introduces several new fields
that belong to certain high dimensional representations of the gauge
group. To induce an effective mass operator one has to go through
several steps to connect those fields to the light lepton fields
which are in low dimensional representations. In this multistep
seesaw, a tiny neutrino mass can be induced without requiring all
new particles to be very heavy or their couplings to light particles
to be all small.

A realistic model in the second approach has been recently proposed
in Ref \cite{Babu:2009aq}. It introduces a vectorlike fermion
triplet and a scalar four-plet so that the effective operator
responsible for a neutrino mass first appears at dimension seven.
The potential signatures of the new particles at the Tevatron and
LHC have been studied with a special focus on the leptonic decays of
the triply charged scalars. The idea of employing higher-dimensional
representations has been further pursued in Ref \cite{Picek:2009is},
where the neutrino mass is induced from a dimension nine operator.
For a systematic effective field theory approach to neutrino mass
operators of a dimension higher than five, we refer to Ref
\cite{Bonnet:2009ej}.

Any mechanism for the generation of a neutrino mass and mixing is
necessarily correlated with the physics of charged leptons. Before
one can be sure that the physical processes involving new heavy
particles at high energy colliders are relevant to neutrino physics,
it is necessary to examine that the parameter regions assumed in the
analysis of high energy processes are respected by precision low
energy tests in the charged leptons. Particularly relevant in this
respect are lepton flavor violating (LFV) decays of charged leptons
and muon-electron ($\mu e$) conversion in nuclei that are severely
suppressed in the standard model (SM). The experimental bounds on
LFV decays of the muon are already very stringent
\cite{Brooks:1999pu, Bellgardt:1987du}, and the sensitivity to its
radiative decay is expected to be upgraded by orders of magnitude in
the MEG experiment within the next few years
\cite{Baracchini:2010dy}. Significant progress has also been made in
LFV decays of the tau lepton, thanks to the large data sample
collected in recent years at the $B$ factories \cite{Aubert:2009tk,
Hayasaka:2007vc, Marchiori:2009ww, Miyazaki:2007zw}. Associated with
the radiative decays of charged leptons are the precise measurements
of or stringent bounds on their electromagnetic dipole moments
\cite{Odom:2006zz, Bennett:2006fi,edm}. For $\mu e$ conversion in
nuclei, the current most stringent constraints arise for titanium
and gold \cite{Dohmen:1993mp,Bertl:2006up}. PRISM/PRIME is expected
to enhance their experimental sensitivity by several orders of
magnitude in the future \cite{Kuno:2005mm}. These bounds will
provide strong constraints on the parameter space that will be
useful in assessing the feasibility of detecting collider processes
relevant to the neutrino mass generation. This motivates us to do a
systematic investigation of the LFV transitions of the charged
leptons in the model of high dimensional representations
\cite{Babu:2009aq}.LFV decays have been previously studied in a
similar fashion in various models of neutrino mass generation, like
supersymmetric models \cite{Hisano:1995cp}, seesaw models
\cite{Kakizaki:2003jk, Abada:2008ea, Abada:2007ux}, mirror fermions
\cite{Hung:2007ez}, little Higgs \cite{Choudhury:2006sq}, and
color-octet particles \cite{FileviezPerez:2009ud}, to mention a few
amongst many. Reader should consult Refs.
\cite{Kuno:1999jp,Masiero:2004js} for a more complete list of
literature on the subject. Similarly, the $\mu e$ conversion in
nuclei has also been widely considered in many scenarios of new
physics beyond SM, such as supersymmetric models
\cite{Brignole:2004ah}, seesaw models \cite{Cheng:1976uq}, littlest
Higgs with T parity \cite{delAguila:2010nv}, $Z^\prime$ models
\cite{Bernabeu:1993ta}, and so on. The formulas and elaborate
discussion of $\mu e$ conversion in nuclei have been given in
Refs.\cite{Kuno:1999jp,Kitano:2002mt,Czarnecki:1998iz}.

In the next section we shall make a complete analysis on the flavor
structure in the model of high dimensional representation. The mass
matrices are parametrized in terms of physical parameters and then
diagonalized approximately. The LFV decays, the contribution to
dipole moments of charged leptons, and $\mu e$ conversion in nuclei
are then calculated in Sec.\ref{sec:analytic}. In Sec.
\ref{sec:numerical} we illustrate our numerical results by sampling
over a few parameters that are potentially interesting. We discuss
and conclude in the last section.

\section{Model}
\label{sec:model}

To avoid the occurrence of the dimension-five neutrino mass operator
at tree level, one should exclude the fields that carry the same
quantum numbers as those in the three canonical seesaw models. Since
the neutrinos are in the doublet representation, the easiest
approach would be to arrange a Yukawa coupling that connects the
neutrinos to a new scalar field and a new fermion field which differ
in weak isospin by $1/2$. One way to accomplish this is to introduce
a scalar multiplet with weak isospin $3/2$ and a fermion multiplet
with weak isospin $1$. This avoids the type 1 and type 2 seesaws
automatically, while the type 3 is avoided by assigning a different
hypercharge to the fermion multiplet. The is indeed the basic idea
behind the model building in Ref. \cite{Babu:2009aq}. The new fields
are denoted as
\begin{eqnarray}
&&\Phi=\left(\begin{array}{l} %
\Phi_3\\\Phi_2\\\Phi_1\\\Phi_0
\end{array}\right)~~(3/2,3/2),~~~
\Sigma=\left(\begin{array}{l} %
\Sigma_2\\\Sigma_1\\\Sigma_0
\end{array}\right)~~(1,1),~
\end{eqnarray}
where the numbers in the parentheses stand for the weak isospin $I$
and hypercharge $Y/2$, respectively, and the subscripts to the
fields indicate the electric charges in units of $|e|$. The fermion
fields $\Sigma$ are assumed to be vectorlike to avoid chiral
anomaly. The relevant SM fields are the Higgs doublet and the lepton
fields (with the subscripts $L,~R$ denoting chirality),
\begin{eqnarray}
&&H=\left(\begin{array}{c}H^+\\H^0\end{array}\right)~~(1/2,1/2),~~~
F_L=\left(\begin{array}{c}n_L\\f_L\end{array}\right)~~(1/2,-1/2),
~~~f_R~~(0,-1).
\end{eqnarray}

\subsection{Yukawa couplings and mass matrices}
\label{subsec:model_1}

The neutrality in the hypercharge allows the following terms and
their Hermitian conjugates: $F_L^*f_RH$, $F_Lf_R\Phi$, $F_L\Sigma
H^*$, $F_L\Sigma^*\Phi$, $\Sigma^*\Sigma$. It is possible to assign
the lepton number, $L(f_R)=L(F_L)=1,~L(\Sigma)=-1,~L(\Phi)=-2$. Then
$L$ is violated when $\Phi$ develops a vacuum expectation value
(VEV). Now we write the terms in a form that respects $SU(2)_L$ and
Lorentz symmetries. The first and last terms are trivial,
$\overline{F_L}f_RH,~\overline{\Sigma}\Sigma$, while the second one
is forbidden. For the third term, $H^*$ should be replaced by
$\tilde H=i\sigma_2H^*$ to preserve its identity as a doublet. To
form a Lorentz scalar without complex conjugation out of two spinor
fields, we can use the charge-conjugated fields. The required
invariant form for the third term is, in terms of Clebsch-Gordan
coefficients,
\begin{eqnarray}
\Big(\overline{F_L^C}\tilde H\Sigma\Big)_0&=&
\frac{1}{\sqrt{3}}\big( -\overline{f_L^C}H^-\Sigma_2
+\overline{n_L^C}H^{0*}\Sigma_0\big)
-\frac{1}{\sqrt{6}}\big(\overline{f_L^C}H^{0*}
-\overline{n_L^C}H^-\big)\Sigma_1,
\end{eqnarray}
where the subscript $0$ on the left indicates its weak isospin of
the product. We have used the notation for charge conjugation that
$\psi_L^C=(\psi_L)^C$ and $\psi^C=\calC\gamma^0\psi^*$ with
$\calC=i\gamma^0\gamma^2$.

For the fourth term in the list we note that, since the vector
representation of $SU(2)$ is strictly real,
$(\Sigma_0^*,\Sigma_1^*,\Sigma_2^*)$ is a vector when
$(\Sigma_2,\Sigma_1,\Sigma_0)$ is. The invariant form is thus,
\begin{eqnarray}
\Big(\overline\Sigma F_L\Phi\Big)_0&=&
\frac{1}{2}\big(\overline{\Sigma_0}n_L\Phi_0%
-\overline{\Sigma_2}f_L\Phi_3\big)
+\frac{1}{2\sqrt{3}}\big(\overline{\Sigma_2}n_L
\Phi_2-\overline{\Sigma_0}f_L\Phi_1\big)\nonumber\\
&&+\frac{1}{\sqrt{6}}\big(\overline{\Sigma_1}f_L
\Phi_2-\overline{\Sigma_1}n_L\Phi_1\big).
\end{eqnarray}
Including the generation index in SM, the mass terms and Yukawa
couplings are summarized as follows:
\begin{eqnarray}
-\calL_{\textrm{Yuk+mass}}&=&m_\Sigma\overline{\Sigma}\Sigma
+\Big[y_{ij}\overline{F_{Li}}f_{Rj}H%
+x_j\Big(\overline{F_{Lj}^C}\tilde H\Sigma\Big)_0%
+z_j\Big(\overline\Sigma F_{Lj}\Phi\Big)_0+\textrm{h.c.}\Big],
\end{eqnarray}
where $y$ is a $3\times 3$ complex matrix, $x,~z$ are each a
three-component complex column vector, and $m_\Sigma$ is real
positive by definition.

When the electric neutral components of $H$ and $\Phi$ develop a
VEV,
\begin{eqnarray}
\langle H^0\rangle=\frac{\rmv_2}{\sqrt{2}},~ \langle
\Phi_0\rangle=\frac{\rmv_4}{\sqrt{2}},
\end{eqnarray}
the Yukawa terms will contribute to the masses of the neutral and
singly charged fermions. For simplicity we shall assume in this work
that both VEV's are real positive. The mass terms are written as
\begin{eqnarray}
\calL_{\textrm{mass}}&=&\calL_{\textrm{mass,2}}+\calL_{\textrm{mass,1}}
+\calL_{\textrm{mass,0}},
\end{eqnarray}
with the number in the subscript denoting the electric charge. The
field $\Sigma_2$ has only a bare mass:
\begin{eqnarray}
-\calL_{\textrm{mass,2}}&=&m_\Sigma\overline{\Sigma_2}\Sigma_2,
\end{eqnarray}
while the fields of other charges also derive masses from Yukawa
couplings so that mixing between the light and heavy particles can
appear. For the singly charged fields, we have
\begin{eqnarray}
-\calL_{\textrm{mass,1}}&=&\overline{\Psi_{1L}}M_1\Psi_{1R}
+\overline{\Psi_{1R}}M_1^\dagger\Psi_{1L},
\end{eqnarray}
where the four-component column fields and the $4\times 4$ mass
matrix are
\begin{eqnarray}
&&\Psi_{1R}=\left(\begin{array}{l}f_R\\
\Sigma_{1L}^C\end{array}\right),~
\Psi_{1L}=\left(\begin{array}{l}f_L\\
\Sigma_{1R}^C\end{array}\right),~M_1=\left(\begin{array}{cc}
\frac{\rmv_2}{\sqrt{2}}y&-\frac{\rmv_2}{2\sqrt{3}}x^*\\
0&m_\Sigma\end{array}\right).
\end{eqnarray}
Since the neutral fermions are generically Majorana, their mass
terms are apparently more complicated. With the help of
charge-conjugated fields, they are
\begin{eqnarray}
-\calL_{\textrm{mass,0}}&=&\frac{1}{2}\overline{\Psi_{0R}}M_0\Psi_{0L}
+\frac{1}{2}\overline{\Psi_{0L}}M_0^\dagger\Psi_{0R},
\end{eqnarray}
where the fields and mass matrix are
\begin{eqnarray}
&&\Psi_{0R}=\left(\begin{array}{l}n_L^C\\
\Sigma_{0L}^C\\\Sigma_{0R}\end{array}\right),~
\Psi_{0L}=\left(\begin{array}{l}n_L\\
\Sigma_{0L}\\\Sigma_{0R}^C\end{array}\right),~%
M_0=\left(\begin{array}{ccc}
0_{3\times 3}&\frac{\rmv_2}{\sqrt{6}}x&\frac{\rmv_4}{2\sqrt{2}}z\\
\frac{\rmv_2}{\sqrt{6}}x^T&0&m_\Sigma\\
\frac{\rmv_4}{2\sqrt{2}}z^T&m_\Sigma&0\end{array}\right).
\end{eqnarray}
These mass matrices will be analyzed and diagonalized in the later
subsection.

\subsection{Scalar potential}
\label{subsec:model_2}

Before we diagonalize the mass matrices of leptons we discuss the
scalar potential for completeness. Remember that
the fields $\Phi$ and $H$ have the quantum numbers $I=Y/2=3/2,~1/2$,
respectively. The possible quadratic terms are, $H^\dagger H$ and
$\Phi^\dagger\Phi$, while there can be no trilinear terms. For the
quartic terms there are two sets of them: either $\Phi$ and
$\Phi^\dagger$, and $H$ and $H^\dagger$ come in pairs, or one
$\Phi^\dagger$ is accompanied by three $H$. For the first set, the
following ones are obvious:
\begin{eqnarray}
(1a)&&(H^\dagger H)^2,~(\Phi^\dagger\Phi)^2,~ (H^\dagger
H)(\Phi^\dagger\Phi).
\end{eqnarray}
The pure $H$ term is unique. This can also be understood as follows:
with two identical $H$'s of $I=1/2$ one can only construct an $I=1$
form that is symmetric in the two $H$'s. (The $I=0$ form vanishes
identically.) This is also the case with two $\tilde H$'s. From the
two $I=1$ forms one can construct a unique $I=0$ term (that is
symmetric, though not necessarily, in the two forms). Indeed, it is
easy to check that $H^\dagger\tau^a HH^\dagger\tau^a H=(H^\dagger
H)^2$ where $\tau^a$ are the Pauli matrices. But this is not the
case with the $\Phi$ field. With two identical $\Phi$'s of $I=3/2$
we can construct two forms that are symmetric in them, one of $I=3$
and the other of $I=1$. This implies that there are two independent,
invariant, pure $\Phi$ terms. Similarly with the half-$H$ and
half-$\Phi$ terms. The additional terms in the first set are thus
\begin{eqnarray}
(1b)&&(\Phi^\dagger T^a_{3/2}\Phi)(\Phi^\dagger T^a_{3/2}\Phi),~
(H^\dagger\tau^aH)(\Phi^\dagger T^a_{3/2}\Phi),
\end{eqnarray}
where $T^a_{3/2}$ stand for the generator matrices for
$I=3/2$.

For the second set of quartic terms involving three $\tilde H$'s and
one $\Phi$, we first combine three $\tilde H$'s into a form of
$I=\frac{3}{2}$ which must be symmetric due to Bose symmetry. Then,
out of this form and one $\Phi$ field, we form an $I=0$ term:
\begin{eqnarray}
\big(\Phi\tilde H\tilde H\tilde H\big)_0&=&
\frac{1}{2}\Phi_0(H^{0*})^3+\frac{\sqrt{3}}{2}\Phi_1H^-(H^{0*})^2
+\frac{\sqrt{3}}{2}\Phi_2(H^-)^2H^{0*}+\frac{1}{2}\Phi_3(H^-)^3.
\end{eqnarray}
The normalization in the above term looks a bit unusual due to the
appearance of identical fields. The most general scalar potential is
thus
\begin{eqnarray}
V&=&-\mu^2_HH^\dagger H-\mu^2_\Phi\Phi^\dagger\Phi +\lam_H(H^\dagger
H)^2+\lam_\Phi(\Phi^\dagger\Phi)^2 +\lam_\Phi'(\Phi^\dagger
T^a_{3/2}\Phi)(\Phi^\dagger T^a_{3/2}\Phi)
\nonumber\\
&&+\lam H^\dagger H\Phi^\dagger\Phi%
+\frac{1}{2}\lam'(H^\dagger\tau^aH)(\Phi^\dagger T^a_{3/2}\Phi)
+\left[\kappa\big(\Phi\tilde H\tilde H\tilde H\big)_0
+\textrm{h.c.}\right],
\end{eqnarray}
where  all couplings except $\kappa$ are real. We note in passing that
the $\lam_\Phi'$ term was missing in Ref. \cite{Babu:2009aq}.

The VEVs of the scalar fields are determined by requiring the
vanishing of the first derivatives and positive-definiteness of the
matrix of the second derivatives of the potential. We can always
choose, by a global $U(1)_Y$ transformation, one of the VEVs, say,
$\rmv_2$, to be real positive. Then one can see that $\kappa\rmv_4$
must be real, and the vanishing conditions become
\begin{eqnarray}
&&-\mu^2_H+\lam_H\rmv_2^2+\frac{1}{2}\lam|\rmv_4|^2
+\frac{3}{8}\lam'|\rmv_4|^2
+\frac{3}{4}\rmv_2\kappa\rmv_4=0,\nonumber\\
&&-\frac{1}{2}\mu^2_\Phi|\rmv_4|^2+\frac{1}{2}\lam_\Phi|\rmv_4|^4
+\frac{9}{8}\lam_\Phi'|\rmv_4|^4+\frac{1}{4}\lam\rmv_2^2|\rmv_4|^2
+\frac{3}{16}\lam'\rmv_2^2|\rmv_4|^2
+\frac{1}{8}\kappa\rmv_4\rmv_2^3=0.
\end{eqnarray}
Since $\rmv_4\ne 0$ breaks the custodial symmetry, it is natural to assume
$|\rmv_4|\ll\rmv_2$. Assuming further that the quartic couplings are
perturbative, we have to good precision that
\begin{eqnarray}
\rmv_2\approx\sqrt{\frac{\mu^2_H}{\lam_H}},
~~~\frac{\rmv_4}{\rmv_2}\approx\frac{2\kappa^*\mu_H^2}
{8\lam_H\mu^2_\Phi-(4\lam+3\lam')\mu_H^2}.
\end{eqnarray}
Since the $\kappa$ term breaks lepton number, it would be easy to attribute the
tininess of $\rmv_4$ to that of $\kappa$.
With so many free parameters at hand it is no problem to guarantee that the
above is the true vacuum. If $\kappa$ vanishes, it would be necessary to fine-tune
the parameters to get a tiny $\rmv_4$, which we shall not pursue further.

\subsection{Diagonalization of lepton mass matrices}
\label{subsec:model_3}

We continue to assume for the sake of simplicity that both VEVs are
real positive. To diagonalize the mass matrices, we first
parameterize them without losing generality in terms of independent
physical parameters by following the procedure advocated in Ref.
\cite{Liao:2006rn}. We sketch below how this is done.

With three generations in SM there is always one massless neutral
mode, which can be $\nu_1$ [in normal hierachy (NH)] or $\nu_3$ [in
inverted hierarchy (IH)] according to the oscillation data. (With
$n_g\ge 3$ generations there are $n_g-2$ massless modes while there
is none with less generations.) We describe the case of NH in some
detail and will record the result for IH later. By applying a
judicious unitary transformation to $n_{Lj}$, we can convert the
column vectors $x$ and $z$ into the standard form:
\begin{eqnarray}
X=(0,0,x)^T,~Z=(0,z,c_z)^T,%
\label{eq_XZ_NH}
\end{eqnarray}
where $x,~z$ are real positive and $c_z$ is generally complex.
This fixes the phases of the two fields (named again as $n_{L2}$ and $n_{L3}$)
orthogonal to the massless mode ($n_{L1}$)
but leaves the latter's phase free.
To keep the partnership under $SU(2)_L$ between the neutral $n_L$ and charged $f_L$
fields generation by generation, the same transformation should be applied to
$f_{Lj}$ as well. This modifies the entries in $y$ but does not alter
its generality. We assume that this has been done already.

To reduce the $y$ matrix to its minimal form, we proceed as follows.
By a unitary transformation of $f_{Rj}$, we can cast $y$ into the form:
\begin{eqnarray}
y=\left(\begin{array}{ccc}%
r_1e^{i\alpha_1}&r_2e^{i\alpha_2}&r_3e^{i\alpha_3}\\
0&y_2&c_2\\0&0&y_3\end{array}\right),
\end{eqnarray}
where $y_2$ and $y_3$ are real positive and $c_2$ is complex. This
fixes the phases of $f_{R3,R2}$ but leaves free that of $f_{R1}$.
By rephasing further $f_{1L}\to e^{i\beta}f_{1L}$, which is augmented with
$n_{1L}\to e^{i\beta}n_{1L}$ to preserve the $SU(2)_L$ partnership,
the first row of $y$ becomes effectively,
$e^{-i\beta}(r_1e^{i\alpha_1},r_2e^{i\alpha_2},r_3e^{i\alpha_3})$.
Now we choose $\beta=\alpha_2$ (or equally well, $\beta=\alpha_3$),
which fixes the phase of $f_{1L}$ and thus that of the massless
$n_{1L}$ as well,
and then choose $\alpha_1=\beta$, which fixes the phase of
$f_{R1}$. This leaves with us the final version of the $y$ matrix:
\begin{eqnarray}
Y=\left(\begin{array}{ccc}%
y_1&y_4&c_1\\0&y_2&c_2\\0&0&y_3\end{array}\right),%
\label{eq_Y_NH}
\end{eqnarray}
where $y_{1,2,3,4}$ are real positive and $c_{1,2}$ are generally complex.

We have used up the degrees of freedom in defining fermion fields to
reduce the number of parameters in the Yukawa sector to its minimum,
i.e., in terms of independent physical parameters, without
sacrificing generality. There are seven real positive parameters
($m_\Sigma$ and $x,~y_{1,2,3,4},~z$) and three complex ones
($c_{1,2,z}$). They will be traded for nine masses (of one doubly
charged, four singly charged, and four neutral fermions), three CP
phases, and a single independent mixing angle. It looks challenging
at first sight for the model to accommodate the two large mixing
angles measured in oscillation. But as we shall see later, all
heavy fermions are nearly degenerate, which effectively saves
parameters at our disposal. The mass matrices for the singly
charged and neutral fermions are summarized as
\begin{eqnarray}
&&M_1=m_\Sigma\left(\begin{array}{cc} \frac{\epsilon_2}{\sqrt{2}}Y
&-\frac{\epsilon_2}{2\sqrt{3}}X\\
0_{1\times 3}&1\end{array}\right),~
~M_0=m_\Sigma\left(\begin{array}{ccc} %
0_{3\times 3}&\frac{\epsilon_2}{\sqrt{6}}X
&\frac{\epsilon_4}{2\sqrt{2}}Z\\
\frac{\epsilon_2}{\sqrt{6}}X^T&0&1\\
\frac{\epsilon_4}{2\sqrt{2}}Z^T&1&0
\end{array}\right),
\end{eqnarray}
where $\epsilon_2=\rmv_2/m_\Sigma, ~\epsilon_4=\rmv_4/m_\Sigma$ with
$\epsilon_4$ being necessarily tiny. The submatrices $X,~Y,~Z$ for
NH have been given in Eqs. (\ref{eq_XZ_NH}) and (\ref{eq_Y_NH}),
while for IH they are
\begin{eqnarray}
&&X=\left(\begin{array}{c} x\\0\\0
\end{array}\right),
~Z=\left(\begin{array}{c} c_z\\z\\0
\end{array}\right),
~Y=\left(\begin{array}{ccc}%
y_1&0&0\\c_1&y_2&0\\c_2&y_4&y_3\end{array}\right).%
\label{eq_XYZ_IH}
\end{eqnarray}

Now we diagonalize the mass matrices $M_0$ and $M_1$ perturbatively
exploiting the hierarchies in parameters. We describe how this is
done for the NH case and will indicate how to obtain the analogous
results for the IH case. To save writing, we put a prime to all
parameters in the standardized Yukawa matrices shown in Eqs.
(\ref{eq_XZ_NH}), (\ref{eq_Y_NH}), and (\ref{eq_XYZ_IH}), and
reserve the unprimed parameters to those that have been multiplied
by a factor of $\epsilon_{2,4}$; namely
\begin{eqnarray}
x=\frac{\epsilon_2}{\sqrt{6}}x',~(z,c_z)=\frac{\epsilon_4}{2\sqrt{2}}(z',c_z'),
~(y_j,c_i)=\frac{\epsilon_2}{\sqrt{2}}(y_j',c_i').
\end{eqnarray}
The symmetric complex matrix $M_0$ is diagonalized to real
nonnegative eigenvalues by a unitary matrix $U$:
\begin{eqnarray}
U^TM_0U=\textrm{diag}(m_1,m_2,m_3,m_4,m_5),
\end{eqnarray}
where $m_1=0$ for NH while $m_3=0$ for IH. $U$ is composed of the five column
vectors $u(j)$ corresponding to the eigenvalues $m_j$, $U=(u(1),\cdots,u(5))$.
Since $\epsilon_4$
is tiny, we can solve the problem perturbatively in the $z,~c_z$ parameters
(collectively denoted as $\epsilon$) while keeping exact dependence on $x$.
After some work, we obtain the eigenvalues for NH:
\begin{eqnarray}
m_1&=&0,\nonumber\\
\frac{m_2}{m_\Sigma}&=&s_0\Big[\sqrt{z^2+c_0^2|c_z|^2}-c_0|c_z|\Big]
+O(\epsilon^3),\nonumber\\
\frac{m_3}{m_\Sigma}&=&s_0\Big[\sqrt{z^2+c_0^2|c_z|^2}+c_0|c_z|\Big]
+O(\epsilon^3),\nonumber\\
\frac{m_4}{m_\Sigma}&=&\sqrt{1+x^2}-|c_z|c_0s_0+\frac{1}{2}c_0\Big[c_0^2z^2
+(1-3c_0^2s_0^2)|c_z|^2\Big]+O(\epsilon^3),\nonumber\\
\frac{m_5}{m_\Sigma}&=&\sqrt{1+x^2}+|c_z|c_0s_0+\frac{1}{2}c_0\Big[c_0^2z^2
+(1-3c_0^2s_0^2)|c_z|^2\Big]+O(\epsilon^3), \label{eq_mass_NH}
\end{eqnarray}
corresponding to the light eigenvectors
\begin{eqnarray}
u(1)&=&(1,0,0,0,0)^T,
\nonumber\\
u(2)&=&e^{-i\beta_+}\left(\begin{array}{l}
0\\
c_m\\
s_m(-c_0e^{-i\alpha_z})\\
s_mc_0\Big( -c_0\sqrt{z^2+c_0^2|c_z|^2}-s_0^2|c_z|\Big)\\
s_ms_0e^{-i\alpha_z}
\end{array}\right)+O(\epsilon^2),
\nonumber\\
u(3)&=&e^{-i\beta_-}\left(\begin{array}{l}
0\\
s_m\\
c_mc_0e^{-i\alpha_z}\\
c_mc_0\Big( -c_0\sqrt{z^2+c_0^2|c_z|^2}+s_0^2|c_z|\Big)\\
c_m(-s_0e^{-i\alpha_z})
\end{array}\right)+O(\epsilon^2),
\label{eq_light_NH}
\end{eqnarray}
and the heavy ones
\begin{eqnarray}
u(4)&=&\frac{e^{-i\beta_-}}{\sqrt{2}}\left(\begin{array}{l}%
0\\
c_0^2z\\
-e^{-i\alpha_z}\Big[s_0-\frac{1}{2}c_0^2(2c_0^2-s_0^2)|c_z|\Big]\\
\Big[1+\frac{1}{2}c_0^2s_0|c_z|\Big]\\
-e^{-i\alpha_z}\Big[c_0-\frac{1}{2}c_0s_0(2s_0^2-c_0^2)|c_z|\Big]
\end{array}\right)+O(\epsilon^2),
\nonumber\\
u(5)&=&\frac{e^{-i\beta_+}}{\sqrt{2}}\left(\begin{array}{l}%
0\\
c_0^2z\\
e^{-i\alpha_z}\Big[s_0+\frac{1}{2}c_0^2(2c_0^2-s_0^2)|c_z|\Big]\\
\Big[1-\frac{1}{2}c_0^2s_0|c_z|\Big]\\
e^{-i\alpha_z}\Big[c_0 +\frac{1}{2}c_0s_0(2s_0^2-c_0^2)|c_z|\Big]
\end{array}\right)+O(\epsilon^2).
\label{eq_heavy_NH}
\end{eqnarray}
Here we have parametrized $c_z=|c_z|e^{i\alpha_z}$ and
$e^{i(\alpha_z+2\beta_\pm)}=\pm 1$. And the triangular functions are
defined as follows:
\begin{eqnarray}
&&s_0=\frac{x}{\sqrt{1+x^2}},~c_0=\frac{1}{\sqrt{1+x^2}};
~s_m=\sqrt{\frac{m_2}{m_2+m_3}}, ~c_m=\sqrt{\frac{m_3}{m_2+m_3}}.
\end{eqnarray}
One can tidy up the $O(\epsilon)$ terms in $u(2)$ and $u(3)$
in terms of mass ratios using
\begin{eqnarray}
c_0\Big[c_0\sqrt{z^2+c_0^2|c_z|^2}+(-)s_0^2|c_z|\Big]
=\frac{1}{2s_0}\frac{1}{m_\Sigma}
\big[m_{3(2)}+(c_0^2-s_0^2)m_{2(3)}\big],
\end{eqnarray}
but no similar simplification occurs for the terms in $u(4)$ and $u(5)$.

We make a few comments on the above result. The light neutrinos gain
a mass of order $m\sim x'(z'\textrm{ or
}|c_z'|)\rmv_2\rmv_4/m_\Sigma$. For a scalar potential with
$\mu_\Phi^2\gg\mu_H^2$, corresponding to $\Phi$ particles much
heavier than the SM Higgs, we have from the previous subsection that
$\rmv_4\sim|\kappa|\rmv_2^3/\mu_\Phi^2$. This yields a neutrino mass
that is triply suppressed by heavy scales as designed in Ref.
\cite{Babu:2009aq}. Second, the two heavy neutrinos are almost
degenerate with a splitting as small as the one in light neutrinos:
\begin{eqnarray}
m_5-m_4\approx m_3-m_2\approx 2c_0s_0|c_z|m_\Sigma.
\label{eq_splitting}
\end{eqnarray}
The significance of this at high energy colliders will be explored
in the future work. Third, the result for the IH case whose Yukawa
coupling matrices are parametrized as in Eq. (\ref{eq_XYZ_IH}) is
obtained by first reshuffling the labels of the light solutions,
$(m_1,m_2,m_3)\to(m_3,m_1,m_2)$ and
$(u(1),u(2),u(3))\to(u(3),u(1),u(2))$, in Eqs. (\ref{eq_mass_NH})
and (\ref{eq_light_NH}), and then interchanging the first and third
rows in all $u(j)$.

Now we diagonalize the mass matrix $M_1$ for the singly charged
leptons by bi-unitary transformations, $U_L^\dagger
M_1U_R=\textrm{diag}(m_e,m_\mu,m_\tau,m_\chi)$, where $\chi$ is the
heavy lepton of charge $-|e|$. Since the matrix $y$ is not related
to new physics, its entries should be naturally small. At this stage
we have no idea on how large the parameter $x$ could be, and thus
leave it free. We therefore diagonalize $M_1$ in two scenarios
according to whether the $x$ parameter (scenario A) or the $y$
parameters (B) are treated perturbatively. As will be clear later,
the $x$ parameter is severely constrained by LFV transitions of the
muon so that both scenarios serve as almost equally good
approximations. We shall describe the diagonalization for the NH
case and indicate at the end how to obtain the results for the IH
case.

In scenario A we do perturbation in $x$. We first solve the zeroth
order eigenvalue problems
\begin{eqnarray}
&&(u^L(e)~u^L(\mu)~u^L(\tau))^\dagger\calM^2_L(u^L(e)~u^L(\mu)~u^L(\tau))
=\textrm{diag}(\lambda_e,\lambda_\mu,\lambda_\tau),
\nonumber\\
&&(u^R(e)~u^R(\mu)~u^R(\tau))^\dagger\calM^2_R(u^R(e)~u^R(\mu)~u^R(\tau))
=\textrm{diag}(\lambda_e,\lambda_\mu,\lambda_\tau),
\end{eqnarray}
for the two $3\times 3$ Hermitian matrices which share the same set of
real positive eigenvalues:
\begin{eqnarray}
\calM^2_L&=&\left(\begin{array}{ccc} %
y_1^2+y_4^2+|c_1|^2&y_2y_4+c_1c_2^*&y_3c_1\\
\textrm{c.c.}&y_2^2+|c_2|^2&y_3c_2\\
\textrm{c.c.}&\textrm{c.c.}&y_3^2\end{array}\right),
\nonumber\\
\calM^2_R&=&\left(\begin{array}{ccc} %
y_1^2&y_1y_4&y_1c_1\\
\textrm{c.c.}&y_2^2+y_4^2&y_2c_2+y_4c_1\\
\textrm{c.c.}&\textrm{c.c.}&|c_1|^2+|c_2|^2+y_3^2\end{array}\right).
\end{eqnarray}
We denote the light charged leptons by the beginning Greek letters
$\alpha$ and $\beta$, and introduce the auxiliary variable and
vectors:
\begin{eqnarray}
\delta_\alpha=(1-\lam_\alpha)^{-1},~\xi^L=-\sum_{\beta}\delta_\beta
u_\tau^{L*}(\beta)u^L(\beta),~ \xi^R=-\sum_{\beta}\delta_\beta
u_\tau^{R*}(\beta)u^R(\beta),
\end{eqnarray}
where the subscript in $u^{L,R}_\alpha(\beta)$ denotes their $\alpha$-th entry.
The eigenvalues are found to be
\begin{eqnarray}
\frac{m^2_\alpha}{m^2_\Sigma}&=&\lam_\alpha
-\frac{x^2}{2}|u_\tau^L(\alpha)|^2\delta_\alpha\lam_\alpha+O(x^3)
\nonumber\\
&=&\lam_\alpha-\frac{x^2}{2}y_3^2|u_\tau^R(\alpha)|^2\delta_\alpha
+O(x^3),~\alpha=e,\mu,\tau\\
\frac{m^2_\chi}{m^2_\Sigma}&=&1+\frac{x^2}{2}\sum_{\alpha}
|u^L_\tau(\alpha)|^2\delta_\alpha+O(x^3)
\nonumber\\
&=&1+\frac{x^2}{2}+\frac{x^2}{2}\sum_{\alpha}
y_3^2|u_\tau^R(\alpha)|^2\delta_\alpha+O(x^3),
\end{eqnarray}
where the two expressions from diagonalization are equivalent using
$y_3^2|u_\tau^R(\alpha)|^2=\lam_\alpha|u_\tau^L(\alpha)|^2$, and the
unitary matrices exact to $O(x)$ are
\begin{eqnarray}
U_L&=&\left(\begin{array}{rrrc}%
u^L(e)&u^L(\mu)&u^L(\tau)&\frac{x}{\sqrt{2}}\xi^L\\
\frac{x}{\sqrt{2}}\delta_eu_\tau^L(e)& %
\frac{x}{\sqrt{2}}\delta_\mu u_\tau^L(\mu)& %
\frac{x}{\sqrt{2}}\delta_\tau u_\tau^L(\tau)&1
\end{array}\right)\textrm{diag}(1,1,1,p_L),\\
U_R&=&\left(\begin{array}{rrrc}%
u^R(e)&u^R(\mu)&u^R(\tau)&\frac{xy_3}{\sqrt{2}}\xi^R\\
\frac{xy_3}{\sqrt{2}}\delta_eu_\tau^R(e)& %
\frac{xy_3}{\sqrt{2}}\delta_\mu u_\tau^R(\mu)& %
\frac{xy_3}{\sqrt{2}}\delta_\tau u_\tau^R(\tau)&1
\end{array}\right)\textrm{diag}(1,1,1,p_R).
\end{eqnarray}
In the above, the $O(x)$ phases $p_{L,R}$ are not fixed completely,
but $p_Rp_L^*$ is fixed by requiring that the eigenvalue, $m_\chi$,
in $U_L^\dagger M_1U_R$ be indeed real positive. The same
arbitrariness also occurs in $u^{L,R}$, but the number of physical
phases in $U_{L,R}$ is restricted to two by the phases appearing in
$M_1$. Considering $\lam_\alpha\ll 1$ in practice, we have to good
precision, $\delta_\alpha=1+O(\lam_\alpha)$, so that
$\xi^{L,R}_\gamma=-\delta_{\tau\gamma}+O(\lam_\alpha)$ from
unitarity of $u^L$ and $u^R$, which simplifies things a bit.

In scenario B we instead treat the parameters $(y_j,c_i)$ (denoted
collectively as $\eta$) perturbatively. By solving first the
eigenvalue problems,
\begin{eqnarray}
&&(u^L(e)~u^L(\mu)~u^L(\tau))^\dagger\bar\calM^2_L(u^L(e)~u^L(\mu)~u^L(\tau))
=\textrm{diag}(\lambda_e,\lambda_\mu,\lambda_\tau),
\nonumber\\
&&(u^R(e)~u^R(\mu)~u^R(\tau))^\dagger\bar\calM^2_R(u^R(e)~u^R(\mu)~u^R(\tau))
=\textrm{diag}(\lambda_e,\lambda_\mu,\lambda_\tau),
\label{eq_lambda_B}
\end{eqnarray}
where
\begin{eqnarray}
\bar\calM_L^2=\calM_L^2|_{y_3\to c_+y_3},~ \bar\calM_R^2
=\calM_R^2|_{y_3\to c_+y_3},
\end{eqnarray}
we find the eigenvalues
\begin{eqnarray}
\frac{m^2_\alpha}{m^2_\Sigma}&=&\lam_\alpha+O(\eta^4),
~\alpha=e,\mu,\tau,\\
\frac{m^2_\chi}{m^2_\Sigma}&=&1+\frac{1}{2}x^2+O(\eta^2),
\end{eqnarray}
and the diagonalization unitary matrices:
\begin{eqnarray}
U_L&=&\left(\begin{array}{rrrc}%
u_e^L(e)&u_e^L(\mu)&u_e^L(\tau) &0\\
u_\mu^L(e)&u_\mu^L(\mu)&u_\mu^L(\tau) &0\\
c_+u_\tau^L(e)&c_+u_\tau^L(\mu)&c_+u_\tau^L(\tau) &-s_+\\
s_+u_\tau^L(e)&s_+u_\tau^L(\mu)&s_+u_\tau^L(\tau) &c_+
\end{array}\right)+O(\eta^2),\\
U_R&=&\left(\begin{array}{rrrc}%
u_e^R(e)&u_e^R(\mu)&u_e^R(\tau) &0\\
u_\mu^R(e)&u_\mu^R(\mu)&u_\mu^R(\tau) &0\\
u_\tau^R(e)&u_\tau^R(\mu)&u_\tau^R(\tau) &-y_3s_+c_+\\
y_3s_+c_+u_\tau^R(e)&y_3s_+c_+u_\tau^R(\mu)&y_3s_+c_+u_\tau^R(\tau)
&1
\end{array}\right)+O(\eta^2).
\end{eqnarray}
The triangular functions in this scenario are
\begin{eqnarray}
s_+=\frac{x/\sqrt{2}}{\sqrt{1+x^2/2}},~c_+=\frac{1}{\sqrt{1+x^2/2}}.
\end{eqnarray}

When all parameters $x$ and $(y_i,c_i)$ are small in magnitude and
treated on the same footing, both scenarios yield an identical
result to the leading order. In both scenarios, the following mass
relations among the heavy $\Sigma$ particles hold:
\begin{eqnarray}
m_{\Sigma_2}=m_\Sigma,~ m_\chi=m_\Sigma\sqrt{1+x^2/2},
~m_{\nu_4}\approx m_{\nu_5}=m_\Sigma\sqrt{1+x^2}.
\label{eq_splitting}
\end{eqnarray}
In our later numerical analysis we shall work with scenario B. The
explicit results displayed above are for the NH case. For the IH
case whose matrices are parametrized as in Eq. (\ref{eq_XYZ_IH})
and without changing the increasing order of the mass eigenvalues,
$U_{L,R}$ are obtained from those for NH by interchanging the
$(1,3)$ rows which are computed with the parameter interchanges
$y_1\leftrightarrow y_3$ and $c_1\leftrightarrow c_2$ made.

\subsection{Couplings of leptons}
\label{subsec:model_4}

The gauge interactions of the leptons are
\begin{eqnarray}
&&\calL_g=g_2\left(j^{+\mu}_WW_{\mu}^++j^{-\mu}_WW_{\mu}^-+J^\mu_ZZ_\mu\right)
+eJ^\mu_{\textrm{em}}A_\mu.
\end{eqnarray}
The currents are written first in terms of weak eigenstates and then grouped
into $\Psi_{1R,1L}$ and $\Psi_{0R,0L}$:
\begin{eqnarray}
j^{+\mu}_W&=&\overline{\Psi_{0L}}w_L^0\gamma^\mu\Psi_{1L}
+\overline{\Psi_{0R}}w_R^0\gamma^\mu\Psi_{1R}
+\overline{\Psi_{1R}}w_R^2\gamma^\mu\Sigma^C_{2L}
+\overline{\Psi_{1L}}w_L^2\gamma^\mu\Sigma^C_{2R},
\nonumber\\
c_WJ^\mu_Z&=&\overline{\Psi_{0L}}z_L^0\gamma^\mu\Psi_{0L}
+\overline{\Psi_{1L}}z_L^1\gamma^\mu\Psi_{1L}
+\overline{\Psi_{1R}}z_R^1\gamma^\mu\Psi_{1R}
-(1-2s_W^2)\overline{\Sigma^C_2}\gamma^\mu\Sigma^C_2,
\nonumber\\
J^\mu_{\text{em}}&=&
-\bar\ell\gamma^\mu\ell-2\overline{\Sigma^C_2}\gamma^\mu\Sigma^C_2,
\end{eqnarray}
where $\ell$ stands for all four leptons of charge $-1$. The coupling
matrices are
\begin{eqnarray}
&&w_L^0=\left(\begin{array}{cc} %
\frac{1}{\sqrt{2}}1_3&0\\
0&0\\
0&-1\end{array}\right),~
w_R^0=\left(\begin{array}{cc} %
0_3&0\\
0&-1\\
0&0\end{array}\right),~
w_L^2=w_R^2=\left(\begin{array}{c} %
0_3\\
-1\end{array}\right),
\nonumber\\
&&z_L^0=\left(\begin{array}{ccc}%
\frac{1}{2}1_3&&\\&-1&\\&&1\end{array}\right),~
z_L^1=\left(\begin{array}{cc}%
(-\frac{1}{2}+s_W^2)1_3&\\&s_W^2\end{array}\right),~z_R^1=s_W^21_4,
\end{eqnarray}
with the usual notation $s_W=\sin\theta_W$ and $c_W=\cos\theta_W$.
In terms of mass eigenstates, $\Psi_{0L}=U\nu_L$,
$\Psi_{0R}=U^*\nu_R=U^*\nu_L^C$, $\Psi_{1L}=U_L\ell_L$, and
$\Psi_{1R}=U_R\ell_R$, the weak currents finally become
\begin{eqnarray}
j^{+\mu}_W&=&\overline{\nu_L}\calW_L^0\gamma^\mu\ell_L
+\overline{\nu_R}\calW_R^0\gamma^\mu\ell_R
+\overline{\ell_R}\calW_R^2\gamma^\mu\Sigma^C_{2L}
+\overline{\ell_L}\calW_L^2\gamma^\mu\Sigma^C_{2R},
\nonumber\\
c_WJ^\mu_Z&=&\overline{\nu_L}\calZ_L^0\gamma^\mu\nu_L
+\overline{\ell_L}\calZ_L^1\gamma^\mu\ell_L
+\overline{\ell_R}\calZ_R^1\gamma^\mu\ell_R
-(1-2s_W^2)\overline{\Sigma^C_2}\gamma^\mu\Sigma^C_2,
\end{eqnarray}
where
\begin{eqnarray}
&&\calW_L^0=U^\dagger w_L^0U_L,~\calW_R^0=U^Tw_R^0U_R,~%
\calW_L^2=U_L^\dagger w_L^2,~\calW_R^2=U_R^\dagger w_R^2,
\nonumber\\
&&\calZ_L^0=U^\dagger z_L^0U,~\calZ_L^1=U^\dagger_Lz_L^1U_L,~
\calZ_R^1=U^\dagger_Rz_R^1U_R=s_W^21_4.
\end{eqnarray}

The upper-left $3\times 3$ submatrix in $\sqrt{2}\calW_L^{0\dagger}$
($\sqrt{2}$ from our normalization convention for currents)
corresponds to the Pontecorvo-Maki-Nakagawa-Sakata (PMNS) matrix in
the seesaw limit, and will be called the effective PMNS matrix or
$\bar V_\textrm{PMNS}$. To help understand the procedure to be
adopted in our later numerical analysis, we take a closer look at
it. As both scenarios A and B are similar for a not large $x$
parameter, we illustrate our discussion in the latter. The submatrix
is
\begin{eqnarray}
\bar V_\textrm{PMNS}&=&U_{fL}^\dagger\textrm{diag}(1,1,c_+/c_0)U_n.
\end{eqnarray}
Here $U_{fL}=(u^L(e)~u^L(\mu)~u^L(\tau))$ is the unitary matrix that
diagonalizes $\bar\calM_L^2$, while $U_n$ is the unitary matrix that
diagonalizes the light neutrino mass in the seesaw limit upon
choosing $e^{-i\beta_+}=e^{i\alpha_z/2}$ and
$e^{-i\beta_-}=ie^{i\alpha_z/2}$. In other words, we have
$V_\textrm{PMNS}=U_{fL}^\dagger U_n$ in the seesaw limit, and
$c_+/c_0\approx 1+x^2/4$ measures the small departure of $\bar
V_\textrm{PMNS}$ from unitarity when the mixing with the heavy
particles is taken into account. As the unitarity has been checked at
a precision not better than a percentage, it is safe if $x$ assumes a value
not larger than, say, $0.1$. This will be fully respected in our
numerical analysis.
This result applies to scenario A as well where $x$ is small by
definition.

We briefly highlight some other features in gauge couplings relevant
to LFV transitions of charged leptons. The light charged leptons
have $O(z,c_z)$ suppressed couplings to heavy neutrinos in
$\calW_L^0$. The massless neutrino decouples from $\calW_R^0$, while
all light charged leptons are suppressed in it by a factor of
$xy_3$. A similar situation occurs in the charged current involving
singly and doubly charged leptons: the left-handed part of the light
charged leptons is suppressed by $x$ and the right-handed by $xy_3$.
Finally, the neutral current of the singly charged leptons is
dominantly flavor diagonal, with an $O(x)$ mixing between the light
and heavy particles and an $O(x^2)$ mixing amongst the light
leptons.

As most new gauge couplings between light and heavy particles are
controlled by the $x$ parameter, we shall consider its largest
allowed value in numerical analysis. The spectrum of the light
neutrinos then implies that the $z$ parameters must be extremely
small. We can therefore focus in the Yukawa sector on the $x$ terms.
Ignoring the tiny mixture between the doublet and four-plet scalars
that is proportional to $\rmv_4/\rmv_2$, the terms relevant to our
study are
\begin{eqnarray}
-\calL_\textrm{Yuk}&\supset&
-\frac{x'h}{2\sqrt{3}}\overline{f_{L3}}\Sigma_1^C +\textrm{h.c.}
=-\frac{x'h}{2\sqrt{3}}(U^*_L)_{3\alpha}(U_R)_{4\beta}
\overline{\ell_{L\alpha}}\ell_{R\beta}h+\textrm{h.c.},
\end{eqnarray}
where the summation is over all singly charged leptons and $h$ is the
physical scalar in $H^0=(h+iG^0)/\sqrt{2}$.

\section{Lepton flavor violating transitions}
\label{sec:analytic}

\subsection{Radiative decays and
electromagnetic dipole moments}

A direct consequence of the neutrino mass and mixing mechanism in
the last section is the lepton flavor violating transitions of the
light charged leptons. We start with the radiative decay
$\ell_\alpha\to\ell_\beta\gamma$, whose amplitude has the dipole
form
\begin{eqnarray}
\calA=\frac{\sqrt{2}eG_F}{(4\pi)^2}\bar u_\beta(p-q)
(h_LP_L+h_RP_R)i\sigma_{\mu\nu}u_\alpha(p)\epsilon^\mu(q)q^\nu,
\end{eqnarray}
where $p,~p-q$ are the momenta of the initial and final leptons,
$q,~\epsilon(q)$ are the photon's momentum and polarization vector,
and $P_{L,R}=(1\mp\gamma_5)/2$. All information on dynamics is
stored in the form factors $h_{L,R}$. The decay width is
\begin{eqnarray}
\Gamma(\ell_\alpha\to\ell_\beta\gamma)=\frac{m_\alpha^3\alpha
G_F^2}{2^9\pi^4} (|h_L|^2+|h_R|^2),
\end{eqnarray}
where we have ignored the mass of $\ell_\beta$ in phase space.

\begin{figure}
\begin{center}
\resizebox{0.9\textwidth}{!}{%
\includegraphics{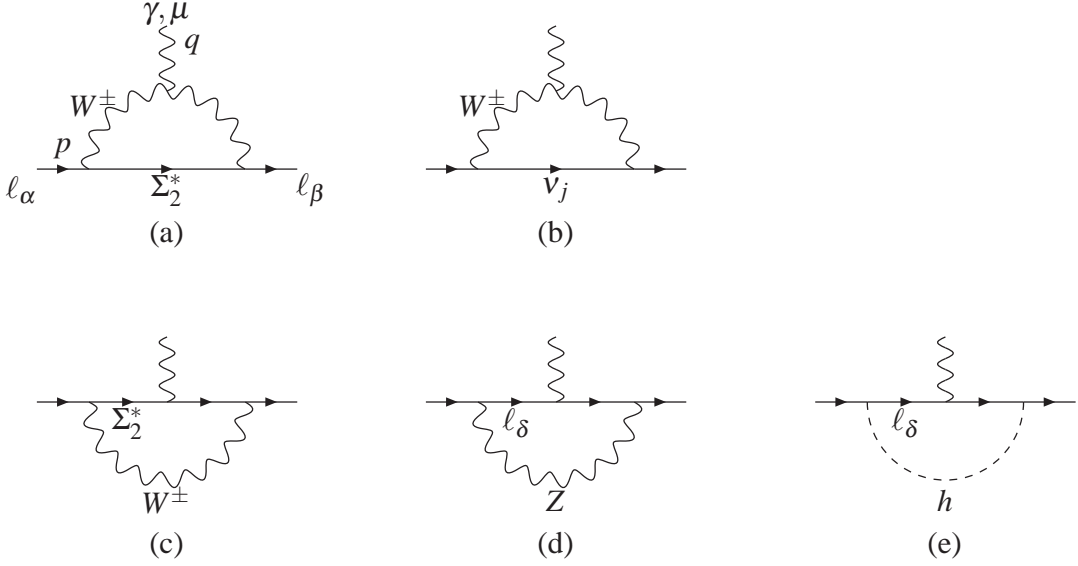}
}
\caption{Feynman diagrams for $\ell_\alpha\to\ell_\beta\gamma$.}%
\label{fig1}
\end{center}
\end{figure}

For the model considered here, the form factors $h_{L,R}$ are
contributed by the Feynman diagrams shown in Fig. \ref{fig1}.
Figures (1a)-(1c) involve the charged currents between the singly
and doubly charged leptons, and between the neutral and singly
charged leptons, respectively, while Figs. (1d) and (1e) originate
from the flavor-changing neutral currents (FCNC) and physical Higgs
exchange. For the gauge-boson mediated graphs we compute in the
unitary gauge. This is simplest but caution must be exercised to
cope with a technical point concerning constant terms, see the last
paper in Ref.\cite{Hung:2007ez}. We have carefully done the Dirac
algebra in $d$ dimensions before the limit $d\to 4$ is taken, to
avoid missing certain finite terms, and find consistent results with
that work.

Ignoring again the terms suppressed by $m_\beta$ and keeping terms
up to the linear order in $m_\alpha$, the form factors from each
graph in Fig. \ref{fig1} are
\begin{eqnarray}
h_L(a)&=&-2\calW_{R\beta}^2\Big[\calW_{R\alpha}^{2*}m_\alpha
\calF(r_\Sigma) +\calW_{L\alpha}^{2*}m_\Sigma \calG(r_\Sigma)\Big],
\nonumber\\
h_L(b)&=&2\calW_{R,j\beta}^{0*}\Big[\calW_{R,j\alpha}^0m_\alpha
\calF(r_j)+\calW_{L,j\alpha}^{0}m_j\calG(r_j)\Big],
\nonumber\\
h_L(c)&=&2\calW_{R\beta}^2\Big[\calW_{R\alpha}^{2*}m_\alpha
\calH(r_\Sigma)+\calW_{L\alpha}^{2*}m_\Sigma\calJ(r_\Sigma)\Big],
\nonumber\\
h_L(d)&=&\calZ^1_{R,\beta\delta}\Big[\calZ^1_{R,\delta\alpha}m_\alpha
\calH(s_\delta)
+\calZ^1_{L,\delta\alpha}m_\delta\calJ(s_\delta)\Big],
\nonumber\\
h_L(e)&=&\frac{1}{\sqrt{2}G_Fm_h^2}\frac{x'^2}{12}U_{R,4\beta}^*U_{L,3\delta}
\Big[U_{L,3\delta}^*U_{R,4\alpha}m_\alpha\calK(t_\delta)
+U_{R,4\delta}^*U_{L,3\alpha}m_\delta\calL(t_\delta)\Big],
\label{eq_hL}
\end{eqnarray}
where summation over the virtual lepton flavors is implied, and
\begin{eqnarray}
h_R(a,b,c,d)=h_L(a,b,c,d)|_{L\leftrightarrow R},
~h_R(e)=h_L(e)|_{L\leftrightarrow R,3\leftrightarrow 4}.
\end{eqnarray}
We have denoted the ratios of the masses appearing in the loops as
$r_\Sigma=m^2_\Sigma/m^2_W$, $r_j=m^2_j/m_W^2$,
$s_\delta=m^2_\delta/m^2_Z$, $t_\delta=m_\delta^2/m_h^2$, where $j$
and $\delta$ enumerate all virtual neutral and singly charged
leptons, respectively. Some products of the coupling matrices in the
above can be simplified using their explicit forms. For instance,
the first term in $h_L(d)$ drops out since $\calZ_R^1$ is diagonal;
the matrices in the charged current involving the doubly charged
lepton are $\calW_{R(L)\alpha}^2=-U^*_{R(L),4\alpha}$. And the loop
functions are
\begin{eqnarray}
\calF(r)&=&\frac{1}{6(1-r)^4}\big[10-43r+78r^2-49r^3+4r^4+18r^3\ln
r\big],
\nonumber\\
\calG(r)&=&\frac{1}{(1-r)^3}\big[-4+15r-12r^2+r^3+6r^2\ln r\big],
\nonumber\\
\calH(r)&=&\frac{1}{3(1-r)^4}\big[-8+38r-39r^2+14r^3-5r^4+18r^2\ln
r\big],
\nonumber\\
\calJ(r)&=&\frac{2}{(1-r)^3}\big[4-3r-r^3+6r\ln r\big],
\nonumber\\
\calK(r)&=&\frac{1}{12(1-r)^4}\big[2+3r-6r^2+r^3+6r\ln r\big],
\nonumber\\
\calL(r)&=&\frac{1}{2(1-r)^3}\big[-3+4r-r^2-2\ln r\big].
\end{eqnarray}

Related to the above radiative transition amplitudes are the
anomalous magnetic moments and electric dipole moments of the singly
charged light leptons. They are worked out to the linear order in
the mass $m_\alpha$ of the considered lepton. The anomalous magnetic
moment, defined as $a=(g-2)/2$, is
\begin{eqnarray}
a(\ell_\alpha)&=&\frac{2\sqrt{2}G_Fm_\alpha}{(4\pi)^2} \Big[
h(a)+h(b)+h(c)+h(d)+h(e)\Big],%
\label{eq_amm}
\end{eqnarray}
where
\begin{eqnarray}
h(a)&=&-2\Big[\big(|\calW_{R\alpha}^2|^2+|\calW_{L\alpha}^2|^2\big)
m_\alpha\calF(r_\Sigma)
+\textrm{Re}\big(\calW_{R\alpha}^2\calW_{L\alpha}^{2*}\big) m_\Sigma
\calG(r_\Sigma)\Big],
\nonumber\\
h(b)&=&2\Big[\big(|\calW_{R,j\alpha}^0|^2+|\calW_{L,j\alpha}^0|^2\big)
m_\alpha\calF(r_j)
+\textrm{Re}\big(\calW_{R,j\alpha}^{0*}\calW_{L,j\alpha}^{0}\big)
m_j\calG(r_j)\Big],
\nonumber\\
h(c)&=&2\Big[\big(|\calW_{R\alpha}^2|^2+|\calW_{L\alpha}^2|^2\big)
m_\alpha\calH(r_\Sigma)
+\textrm{Re}\big(\calW_{R\alpha}^2\calW_{L\alpha}^{2*}\big)m_\Sigma
\calJ(r_\Sigma)\Big],
\nonumber\\
h(d)&=&\big(|\calZ^1_{R,\alpha\delta}|^2+|\calZ^1_{L,\alpha\delta}|^2\big)
m_\alpha\calH(s_\delta)
+\textrm{Re}\big(\calZ^1_{R,\alpha\delta}\calZ^1_{L,\delta\alpha}\big)
m_\delta\calJ(s_\delta),
\nonumber\\
h(e)&=&\frac{1}{\sqrt{2}G_Fm_h^2}\frac{x'^2}{12}\Big[
\big(|U_{L,3\delta}U_{R,4\alpha}|^2+|U_{R,4\delta}U_{L,3\alpha}|^2\big)
m_\alpha\calK(t_\delta)\nonumber\\
&&+\textrm{Re}\big(U_{R,4\alpha}^*U_{L,3\delta}U_{R,4\delta}^*
U_{L,3\alpha}\big)m_\delta\calL(t_\delta)\Big].
\end{eqnarray}
Again some of the above can be simplified using explicit forms of
the coupling matrices given in the last section.

The electric dipole moment of the fermion $\psi$ is defined as
$\calL_\textrm{edm}=-id/2\bar\psi\sigma_{\mu\nu}\gamma_5\psi
F^{\mu\nu}$, with $F^{\mu\nu}$ being the electromagnetic field
tensor. It is evaluated to be
\begin{eqnarray}
d(\ell_\alpha)&=&-\frac{2\sqrt{2}eG_F}{(4\pi)^2}\Big[ \bar h(a)+\bar
h(b)+\bar h(c)+\bar h(d)+\bar h(e)\Big],%
\label{eq_edm}
\end{eqnarray}
where
\begin{eqnarray}
\bar
h(a)&=&-2\textrm{Im}\big(\calW_{R\alpha}^2\calW_{L\alpha}^{2*}\big)
m_\Sigma\calG(r_\Sigma),
\nonumber\\
\bar h(b)&=&
2\textrm{Im}\big(\calW_{R,j\alpha}^{0*}\calW_{L,j\alpha}^{0}\big)
m_j\calG(r_j),
\nonumber\\
\bar h(c)&=&
2\textrm{Im}\big(\calW_{R\alpha}^2\calW_{L\alpha}^{2*}\big) m_\Sigma
\calJ(r_\Sigma),
\nonumber\\
\bar h(d)&=&\textrm{Im}\big(
\calZ^1_{R,\alpha\delta}\calZ^1_{L,\delta\alpha}\big)
m_\delta\calJ(s_\delta),
\nonumber\\
\bar h(e)&=&\frac{1}{\sqrt{2}G_Fm_h^2}
\frac{x'^2}{12}\textrm{Im}\big(U_{R,4\alpha}^*U_{L,3\delta}
U_{R,4\delta}^*U_{L,3\alpha}\big)m_\delta\calL(t_\delta).
\end{eqnarray}
Since $\calZ^1_R$ is diagonal and $\calZ_L$ is Hermitian, there is
actually no contribution from the FCNC graph, $\bar h(d)=0$.

\subsection{Purely leptonic transitions}

Now we consider the purely leptonic transitions of the light charged
leptons. These include the experimentally well-bounded decays
$\ell_\delta\to\ell_\alpha\ell_\beta\bar\ell_\gamma$ and the
muon-electron ($\mu e$) conversion in nuclei. The leading
contributions in the model considered here arise from FCNC couplings
of the $Z$ boson. The Higgs exchange is suppressed by additional
factors of $x$ and a heavier Higgs mass, while we have verified that
the photonic contribution is always subdominant. We do not consider
LFV decays of the $Z$ boson since they are experimentally much less
constrained.

\begin{figure}
\begin{center}
\resizebox{0.5\textwidth}{!}{%
\includegraphics{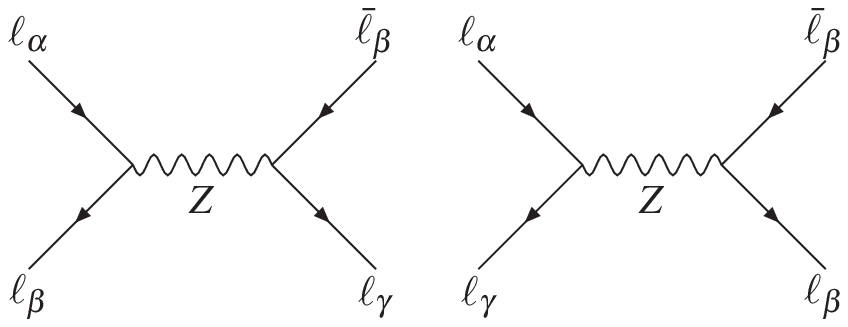}
} \caption{Feynman diagrams for
$\ell_\alpha\to\ell_\beta\ell_\gamma\bar\ell_\beta$.}
\label{fig2}%
\end{center}
\end{figure}

There are three types of decays corresponding to
$\alpha=\beta=\gamma$, $\alpha=\gamma\ne\beta$, and
$\alpha=\beta\ne\gamma$. Since the flavor-changing couplings carry a
factor of $x^2$, only the transitions of the first two types are
important while the third one is severely suppressed. We thus
concentrate on the decays,
$\ell_\alpha\to\ell_\beta\ell_\gamma\bar\ell_\beta$ with
$\gamma=\beta$ or $\gamma\ne\beta$, whose leading terms come from
diagrams shown in Fig. \ref{fig2}. Note that there is a relative
minus sign between the two graphs and that for $\gamma=\beta$ one
should attach a factor of $1/2$ in the total decay rate. Ignoring
the final-state masses, the rate is given by
\begin{eqnarray}
\frac{\Gamma(\ell_\alpha\to\ell_\beta\ell_\gamma\bar\ell_\beta)}
{4\Gamma_0}&=&\frac{1}{1+\delta_{\beta\gamma}}\Big[%
|\calZ_{L,\gamma\alpha}^1\calZ_{L,\beta\beta}^1|^2
+|\calZ_{L,\beta\alpha}^1\calZ_{L,\gamma\beta}^1|^2
\nonumber\\
&&+2\textrm{Re}\big(\calZ_{L,\gamma\alpha}^1\calZ_{L,\beta\beta}^1
\calZ_{L,\beta\alpha}^{1*}\calZ_{L,\gamma\beta}^{1*}\big)
\nonumber\\
&&+|\calZ_{L,\gamma\alpha}^1\calZ_{R,\beta\beta}^1|^2
+|\calZ_{L,\beta\alpha}^1\calZ_{R,\gamma\beta}^1|^2
+(L\leftrightarrow R)\Big]%
\label{eq_rate_3l}
\end{eqnarray}
with $\Gamma_0=G_F^2m_\alpha^5/(192\pi^3)$ being the decay rate for
the dominant decay mode,
$\ell_\alpha\to\nu_\alpha\ell_\beta\bar\nu_\beta$. Since $\calZ^1_R$
is diagonal and $\alpha\ne\beta,~\alpha\ne\gamma$, the term
$(L\leftrightarrow R)$ actually drops out.

A competitive process is the coherent $\mu e$ conversion in nuclei,
$\mu^-N\rightarrow e^-N$. It involves various atomic and nuclear
effects in addition to the short-distance physics of lepton flavor
violation. A comprehensive study has been given in Ref.
\cite{Kitano:2002mt} based on the method developed in Ref.
\cite{Czarnecki:1998iz}, which improved over earlier efforts on
various corrections \cite{Shanker:1979ap,Chiang:1993xz} to the
original calculations \cite{Weinberg:1959zz,Marciano:1977cj}. These
corrections turn out to be particularly important for heavy nuclei.

The effective Lagrangian relevant to the coherent conversion via the
leptonic FCNC couplings of the $Z$ boson can be written as
\cite{Kitano:2002mt}
\begin{eqnarray}
\calL_{\mu e}=-\frac{G_F}{\sqrt{2}}\sum_{q=u,d,s}
\Big[\big(g_{LV(q)}\bar{e}\gamma^\mu P_L\mu
+g_{RV(q)}\bar{e}\gamma^\mu P_R\mu\big)\bar{q}\gamma_\mu
q+\textrm{h.c.}\Big]
\end{eqnarray}
where $g_{LV(u)}=(2-16s^2_W/3)\calZ^1_{L,e\mu},
~g_{LV(d,s)}=(-2+8s^2_W/3)\calZ^1_{L,e\mu}$, and $g_{RV(q)}=0$ for
the considered model. Then, the $\mu e$ conversion rate is
\begin{eqnarray}
\Gamma(\mu^-N\rightarrow e^-N)
=2G^2_F\Big[\big|\widetilde{g}^{(p)}_{LV}V^{(p)}_N+
\widetilde{g}^{(n)}_{LV}V^{(n)}_N\big|^2+
\big|\widetilde{g}^{(p)}_{RV}V^{(p)}_N+
\widetilde{g}^{(n)}_{RV}V^{(n)}_N\big|^2\Big],
\end{eqnarray}
where $\widetilde{g}^{(p)}_{LV}=2g_{LV(u)}+g_{LV(d)},
~\widetilde{g}^{(n)}_{LV}=g_{LV(u)}+2g_{LV(d)}$, and similarly for
$\widetilde{g}^{(p)}_{RV}$ and $\widetilde{g}^{(p)}_{RV}$.
$V^{(p)}_N$ and $V^{(n)}_N$ are overlap integrals of the $\mu,~e$
with the protons and neutrons in the nucleus $N$, which have been
numerically evaluated and cataloged in \cite{Kitano:2002mt}. The
above rate is usually normalized to the corresponding ordinary
muon capture rate, $\omega_\textrm{capt}$, to yield a branching
ratio, $\Br(\mu^-N\rightarrow e^-N)$, for the $\mu e$ conversion
on a particular nucleus $N$. Since the purely leptonic decays and
the $\mu e$ conversion in nuclei originate from the same FCNC
couplings, the ratio of their branching ratios has the simple
form,
\begin{eqnarray}
\frac{\Br(\mu^-N\rightarrow e^-N)}{\Br(\mu\rightarrow ee\bar{e})}=
\frac{G^2_F[(2-8s^2_W)V^{(p)}_N-2V^{(n)}_N]^2}
{\omega_\textrm{capt}(1-4s^2_W+6s^4_W)},
\end{eqnarray}
upon ignoring minor corrections in the diagonal element
$\calZ^1_{L,ee}$. Namely, the relative importance of the two
transitions rests on that of their experimental bounds.

\section{Numerical analysis}
\label{sec:numerical}

As we shall see later, the Yukawa coupling $x'$, or
$x=x'\rmv_2/(\sqrt{6}m_\Sigma)$, that couples the light and heavy
fermions via the Higgs doublet is a central parameter that controls
the overall scale of the LFV transition rates. We mentioned earlier
that the parameter measures the unitarity violation in the effective
PMNS matrix. Since the heavy fermions have a squared mass splitting
proportional to $x^2$ [see Eq.(\ref{eq_splitting})], it could also
be sensitive to the violation of custodial symmetry measured by the
parameter $\Delta\rho=m_W^2/(c_W^2m_Z^2)-1$. We have calculated the
one-loop contribution due to the heavy fermions
\begin{eqnarray}
\Delta\rho_\Sigma=\frac{4\sqrt{2}G_Fm_\Sigma^2}{(4\pi)^2}
\frac{19}{48}x^4\approx 1.7\times 10^{-5}x'^4
\frac{1~\TeV^2}{m_\Sigma^2}.
\end{eqnarray}
This is balanced by that of a nonvanishing VEV of the nondoublet
scalar field, $\rmv_4\ne 0$, that occurs already at the tree level,
$\Delta\rho_\Phi\approx -6\rmv_4^2/\rmv_2^2$ (see also Ref.
\cite{Babu:2009aq}). We noted in the above that half of the power in
$x^4$ comes from the mass splitting of the heavy fermions. The other
half originates from the mixing effect in the two vertices, which is
essential for a contribution to $\Delta\rho$ since a vectorlike
multiplet cannot contribute even if it is not degenerate. Since the
$\rho$ parameter is measured at a precision not better than
$10^{-4}$, we are on the safe side if $x'$ is not larger than $0.7$
even for a doubly charged fermion as light as $200~\GeV$. This is a
much weaker constraint than we shall get below from LFV transitions.

Before we show our numerical results, we outline how the free
parameters are manipulated based upon the formulas in Sec.
\ref{sec:analytic}. Since the unitarity of the PMNS matrix has been
verified to certain level, we can use it as a guide in browsing the
parameter space. In this way we can cover the majority of the
parameter space that is consistent with an almost unitary effective
PMNS matrix, $\bar V_\textrm{PMNS}$. The PMNS matrix generally
contains three mixing angles, one Dirac phase, and two Majorana
phases in the standard form:
\begin{eqnarray}
V_\textrm{PMNS}=\left(\begin{array}{ccc}
c_{12}c_{13}&s_{12}c_{13}&s_{13}u_\delta^*\\
-c_{23}s_{12}-s_{13}s_{23}c_{12}u_\delta&
c_{23}c_{12}-s_{13}s_{23}s_{12}u_\delta& s_{23}c_{13}\\
s_{23}s_{12}-s_{13}c_{23}c_{12}u_\delta&
-s_{23}c_{12}-s_{13}c_{23}s_{12}u_\delta& c_{23}c_{13}
\end{array}\right)\textrm{diag}(u_1,u_2,u_3),
\end{eqnarray}
where $c_{ij}=\cos\theta_{ij}$, $s_{ij}=\sin\theta_{ij}$,
$u_j=\exp(i\alpha_j/2)$, and $u_\delta=\exp(i\delta)$. We use the
measured values for those angles (while choosing some values for the
phases which are not yet measured) and for the light neutrino masses
$m_{1,2,3}$ in either NH or IH. Then, the heavy neutrino masses
$m_{4,5}$ and the parameters $z,~|c_z|$ are uniquely fixed once the
parameters $(x,m_\Sigma)$ are assigned a value. The diagonalization
matrix $U$ for the neutrinos is also fixed up to the phase of
$c_z=|c_z|e^{i\alpha_z}$. In particular, the matrix $U_n$ that is
formed from the first three rows of the vectors in Eq.
(\ref{eq_light_NH}) (for the NH case and similarly for IH) is fixed
up to a diagonal phase matrix,
$U_z\equiv\textrm{diag}(1,e^{i\alpha_z/2},e^{-i\alpha_z/2})$,
multiplied from the left. Applying the definition
$V_\textrm{PMNS}=U_{fL}^\dagger U_n$ and the first equation in Eq.
(\ref{eq_lambda_B}) (for scenario B and similarly for scenario A),
we have $U_z^\dagger\bar\calM_L^2U_z=(U_z^\dagger U_n)
V_\textrm{PMNS} \textrm{diag}(\lambda_e,\lambda_\mu,\lambda_\tau)
V^\dagger_\textrm{PMNS}(U^\dagger_nU_z)$ where the right-hand side
is completely known with the additional input of the light charged
lepton masses. We observe that this procedure then determines
uniquely all of the parameters $y_{1,2,3,4},~c_{1,2}$ as well as
$\alpha_z$ on the left-hand side. The diagonalization matrix $U_L$
is thus fixed. Once the $Y$ matrix is known, we can follow the
formalism in Sec. \ref{sec:analytic} to find the other
diagonalization matrix $U_R$. In a final step we go back to check
that the effective PMNS matrix $\bar V_\textrm{PMNS}$ obtained from
the matrices $U$ and $U_L$ determined above does not violate
unitarity beyond the allowed level.

\begin{figure}
\begin{center}
\resizebox{0.8\textwidth}{!}{%
\includegraphics{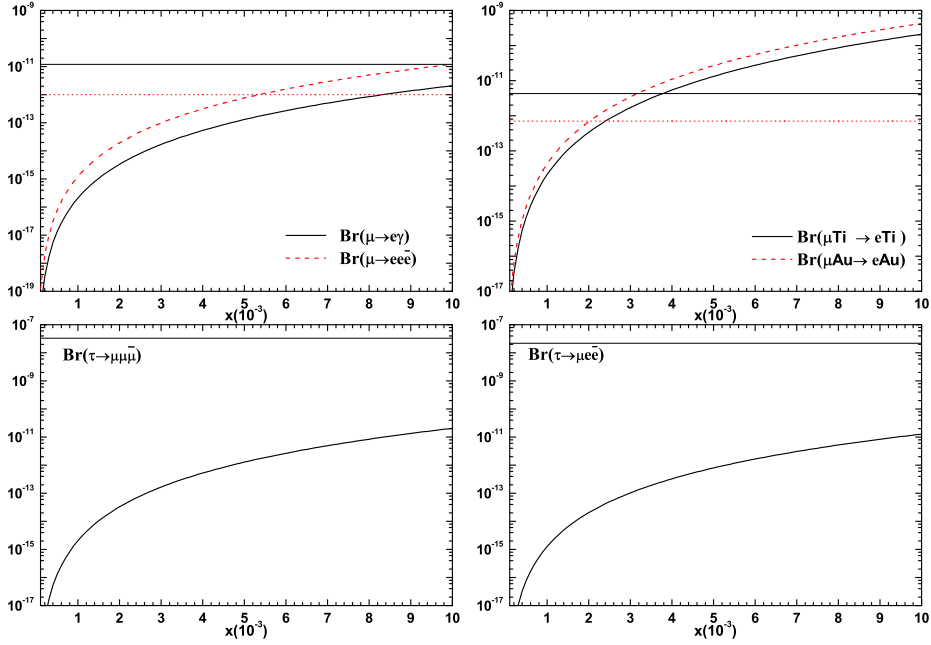}
}%
\caption{Branching ratios as a function of $x$ for NH in scenario B
and tribimaximal mixing.}
\label{fig3}%
\end{center}
\end{figure}

\begin{figure}
\begin{center}
\resizebox{0.8\textwidth}{!}{%
\includegraphics{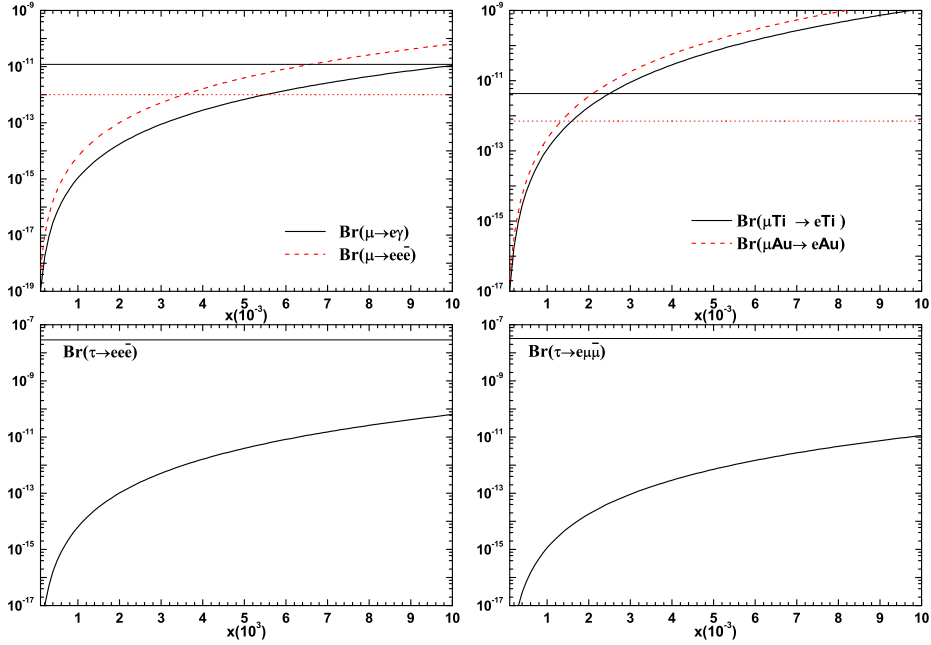}
}%
\caption{Branching ratios as a function of $x$ for IH in scenario B
and tribimaximal mixing.}
\label{fig4}%
\end{center}
\end{figure}

To get some feel about the branching ratios for the LFV transitions,
we start with the results for the simplified case of tribimaximal
neutrino mixing with all phases set to zero. We assume
$m_\Sigma=200~\GeV$ in the following discussions so that the heavy
fermions are within the reach of LHC. In the upper panel of Fig.
\ref{fig3} we show the branching ratios for the muon decays and $\mu
e$ conversion in nuclei $^{197}_{79}\textrm{Au}$ and
$^{48}_{22}\textrm{Ti}$ as a function of the $x$ parameter for the
NH case in scenario B, together with the current experimental bounds
on them (horizontal lines). The lower panel depicts the branching
ratios for the two $\tau$ decays, $\tau\to\mu\mu\bar\mu,~\mu e\bar
e$, for the same range of $x$, while other decays are severely
suppressed. We see that the $\mu e$ conversion on the heavy gold
nucleus sets the most stringent constraint on the $x$ parameter
though it also inherits the largest uncertainty from nuclear
physics. The bound from the conversion on titanium is comparable to
that from the purely leptonic decay, $\mu\to 3e$. For such a small
$x$ the deviation from the SM values of the anomalous magnetic
moments in Eq. (\ref{eq_amm}) and the contribution to the electric
dipole moments in Eq. (\ref{eq_edm}) are too small to be relevant.
In Fig. \ref{fig4} we depict the corresponding results for the IH
case in scenario B. Note that in this case the dominant LFV $\tau$
decays change to $ee\bar e$ and $e\mu\bar\mu$. Generally speaking,
when assuming tribimaximal neutrino mixing, the stringent bounds on
the muon imply that the tau decays are not likely to be observable
in the near future for the majority of the parameter space.

\begin{figure}
\begin{center}
\resizebox{0.8\textwidth}{!}{%
\includegraphics{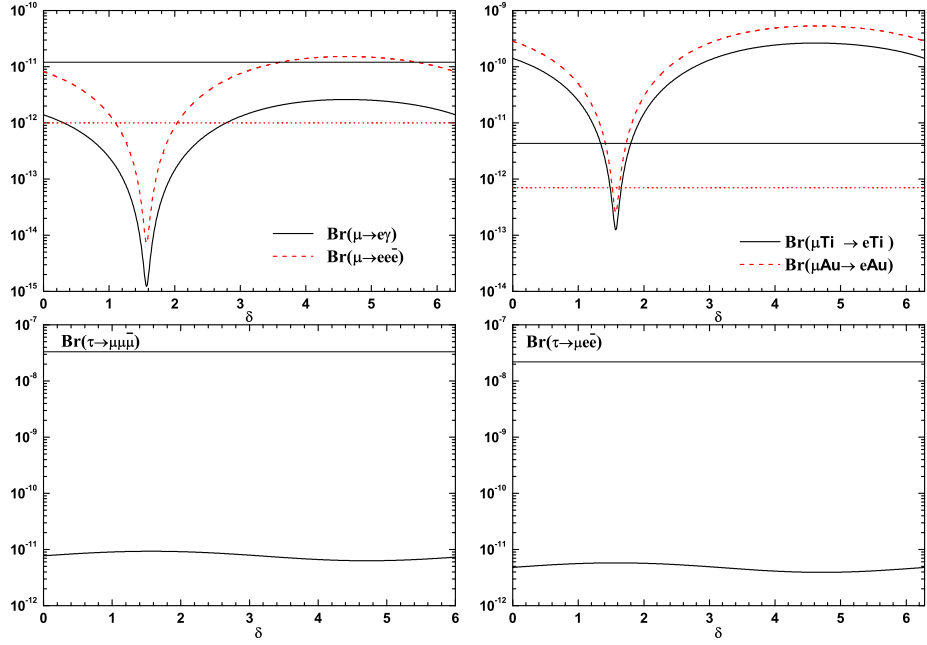}
}%
\caption{Branching ratios as a function of $\delta$ for NH in
scenario B and best-fit mixing angles.}
\label{fig5}%
\end{center}
\end{figure}

\begin{figure}
\begin{center}
\resizebox{0.8\textwidth}{!}{%
\includegraphics{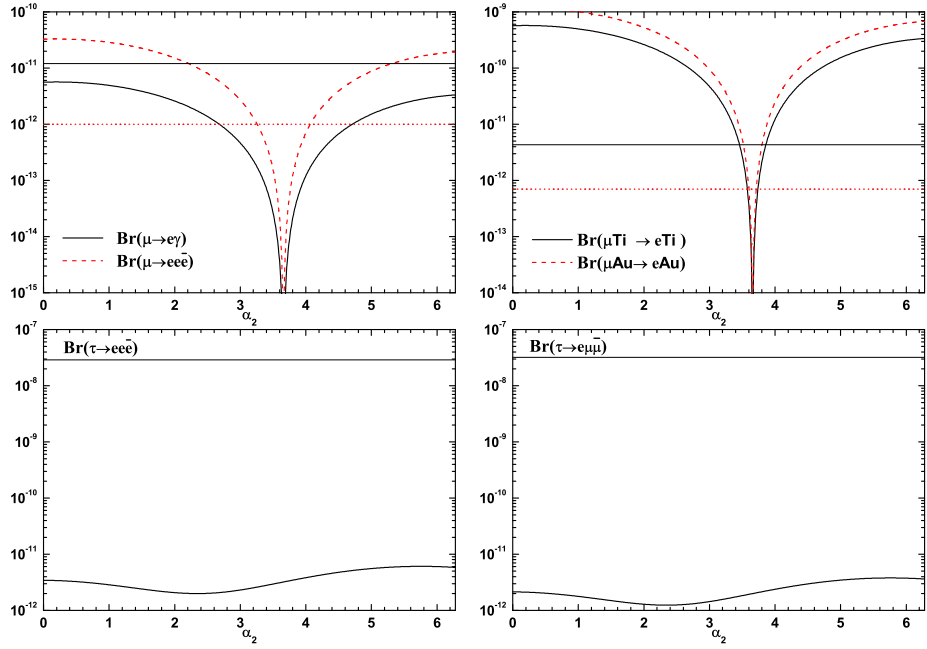}
}%
\caption{Branching ratios as a function of $\alpha_2$ for IH in
scenario B and best-fit mixing angles.}
\label{fig6}%
\end{center}
\end{figure}

We thus ask if there is a region in the parameter space where the
muon decays and $\mu e$ conversion in nuclei are significantly
suppressed while the tau decays are not much below the current
bounds. To help identify the interested region it is useful to work
with the leading terms in the limits of infinite virtual heavy
masses and vanishing virtual light masses. An inspection of Eqs.
(\ref{eq_hL}) and (\ref{eq_rate_3l}) augmented with the coupling
matrices displayed in Sec. \ref{sec:model} tells us that in the NH
case the radiative decay $\ell_\alpha\to\ell_\beta\gamma$ is
dominated by the terms with the mixing matrices
$u^{L*}_\tau(\ell_\beta)u^L_\tau(\ell_\alpha)$,
$u^{R*}_\tau(\ell_\beta)u^L_\tau(\ell_\alpha)$, and
$u^{L*}_\tau(\ell_\beta)u^R_\tau(\ell_\alpha)$, whose coefficients
are less suppressed by a small $x$ parameter, and that the rates for
$\mu\to ee\bar e$ and $\mu e$ conversion in nuclei are proportional
to $|u^{L*}_\tau(e)u^L_\tau(\mu)|^2$. Since for instance in scenario
A,
$|u_\tau^R(\ell_\alpha)|=|u^L_\tau(\ell_\alpha)|\sqrt{\lambda_\alpha}/y_3$
(attach a factor of $c_+$ to $y_3$ for scenario B), we see that the
dominant terms for the LFV muon decays and $\mu e$ conversion in
nuclei are controlled by the combination $\xi_{e\mu}\equiv
u^{L*}_\tau(e)u^L_\tau(\mu)$ [and similarly in the IH case by
$\xi_{e\mu}\equiv u^{L*}_e(e)u^L_e(\mu)$]. We therefore seek for
regions in which the combination $\xi_{e\mu}$ would be significantly
diminished. For the mixing angles and the neutrino squared mass
differences we use the central values from the global fit in Ref.
\cite{Maltoni:2004ei}: $\sin^2\theta_{12}=0.32$,
$\sin^2\theta_{23}=0.50$, $\Delta m^2_{21}=7.6\times 10^{-5}~\eV^2$,
$|\Delta m^2_{31}|=2.4\times 10^{-3}~\eV^2$, and set $\theta_{13}$
at its upper limit, $\sin^2\theta_{13}=0.05$. We choose
$\alpha_1=\alpha_3=0$ while leaving the Dirac phase $\delta$ and
Majorana phase $\alpha_2$ free. For the SM parameters we use the
numbers of the Particle Data Group. We find that $\xi_{e\mu}$
approaches its minimum at $(\delta,\alpha_2)\sim(\pi/2,2\pi)$ for
the NH case and at $(\delta,\alpha_2)\sim(0.64,3.66)$ for IH. This
result is independent of the $x$ parameter (for $x$ not too large in
scenario B).

\begin{figure}
\begin{center}
\resizebox{0.8\textwidth}{!}{%
\includegraphics{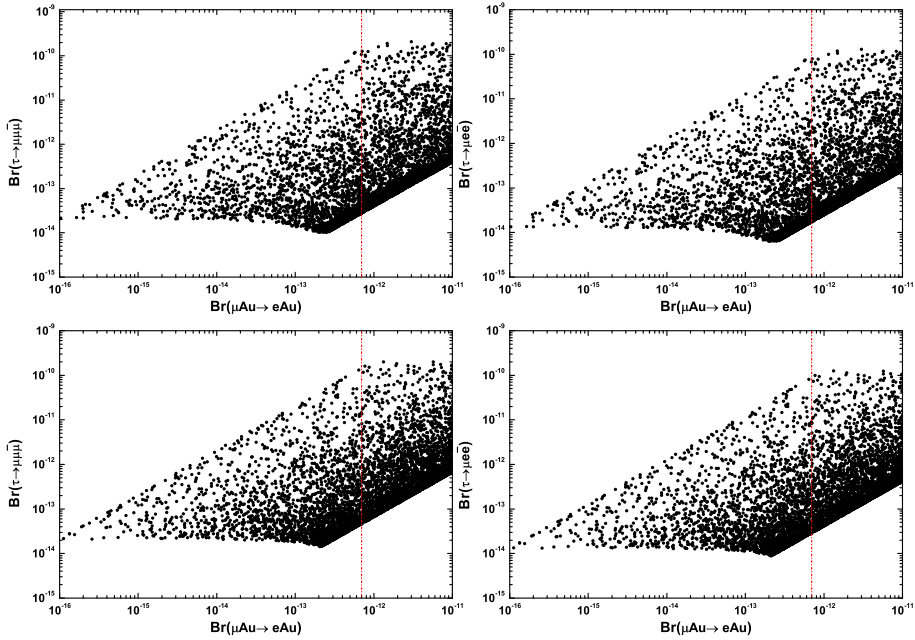}
}%
\caption{Branching ratios sampled over $(x,\alpha_2)$ (upper panels)
and $(x,\delta)$ (lower panels) for NH in scenario B and best-fit
mixing angles.}
\label{fig7}%
\end{center}
\end{figure}

\begin{figure}
\begin{center}
\resizebox{0.8\textwidth}{!}{%
\includegraphics{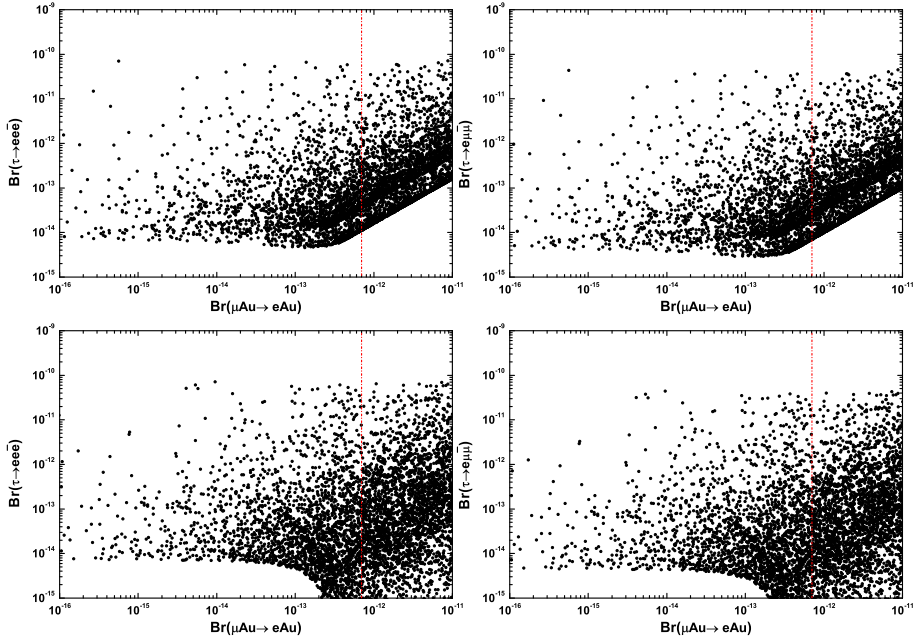}
}%
\caption{Branching ratios sampled over $(x,\alpha_2)$ (upper panels)
and $(x,\delta)$ (lower panels) for IH in scenario B and best-fit
mixing angles.}
\label{fig8}%
\end{center}
\end{figure}

In Fig. \ref{fig5} we show the branching ratios of the muon decays
and $\mu e$ conversion in nuclei and the largest tau decays in the
NH case by scanning over the Dirac phase $\delta$. We have assumed
$x=8\times 10^{-3}$ and $\alpha_2=2\pi$. The corresponding result
for the IH case is depicted in Fig. \ref{fig6} as a function of
$\alpha_2$ at the same $x$ parameter and $\delta=0.64$. One sees
from these two figures that without breaking the stringent bounds on
the muon lepton the branching ratios for some of the leptonic tau
decays can approach the level of $10^{-11}$ for almost the whole
range of the scanned phase. This is much enhanced compared to the
case of tribimaximal mixing shown in Figs. \ref{fig3} and
\ref{fig4}, but is still 3 orders of magnitude below the current
sensitivity.

The above tendency encourages us to scan over a larger set of
parameters. So we finally sample over the $x$ parameter from
$8\times 10^{-4}$ to $8\times 10^{-3}$ and one of the phases in its
whole range while keeping the other phase fixed at the value that
minimizes $\xi_{e\mu}$. Our results are shown in Fig. \ref{fig7} for
the NH and in Fig. \ref{fig8} for the IH case respectively. In both
figures, the upper panel scans over $x$ and the Majorana phase
$\alpha_2$ while the lower one is over $x$ and the Dirac phase
$\delta$. We include only the most stringent $\mu e$ conversion on
gold and the largest tau decay that is available in each case. In
the most optimistic situation, some $\tau$ decays can reach the
level that is about 2 orders of magnitude below the current
sensitivity. We notice that the two figures in the same panel have a
similar pattern. This arises because the decay
$\tau\to\ell_\alpha\ell_\beta\bar\ell_\beta$ with $\alpha\ne\beta$
is dominated by one Feynman graph which is almost the same as any of
the two graphs for the decay
$\tau\to\ell_\alpha\ell_\alpha\bar\ell_\alpha$.

\section{Conclusion}

The origin of tiny neutrino mass has remained mysterious after years
of endeavor. From the viewpoint of effective field theory the tiny
mass can be accommodated by the canonical seesaw mechanisms. But it
is generally hard to explore in current experiments the physics that
would be responsible for the mechanisms because the relevant physics
scale is very high and the new interactions with light particles are
generally too weak. It is thus highly desirable if there is any new
mechanism that would predict accessible effects beyond the neutrino
mass.

There are two basic approaches to relax the tension between the
accessibility of new physics and the effectiveness in producing tiny
neutrino mass. One can either attribute the mass to a higher order
quantum effect or postpone its appearance to a higher-dimensional
effective interaction. An explicit model has been recently attempted
in the second approach \cite{Babu:2009aq}. The idea is to avoid the
conventional dimension-five interaction by composing new fields in
higher-dimensional representations so that the first contribution to
the neutrino mass occurs at dimension seven. The new particles enjoy
the SM gauge interactions, and thus if not very heavy would be
produced at high energy colliders like Tevatron and LHC. The point
that we want to emphasize here is that to establish the kinship of
those particles to the origin of neutrino mass it would be necessary
to detect their interactions with light leptons. These interactions
are as usual shaped by the mixing effects between the light and
heavy particles, and thus should also leave their fingerprints in
precisely measured flavor-changing processes at low energy. The
purpose of the current work has been to examine if there is any
chance to look for the mixing effects in LFV transitions of the
charged leptons.

We have made a systematic analysis of the model. In particular, we
provided a convenient parametrization of the leptons' mass matrices
in terms of independent physical parameters. By diagonalizing them
explicitly the lepton flavor structure becomes transparent in
interactions. The contributions of these interactions to the
radiative, purely leptonic decays of the charged leptons, and the
$\mu e$ conversion in nuclei are then computed. We considered how
the stringent constraints from the muon lepton affect the decay
processes in the tau sector. Generally the current experimental
bounds on the decay $\mu\to ee\bar e$ and the $\mu e$ conversion in
nuclei, in particular are so strong that it is very difficult to
observe the tau lepton decays. However, our sampling over the
unknown phases and Yukawa coupling shows that there are small
regions in the parameter space in which some tau decays have a
branching ratio that is about 2 orders of magnitude below the
current bounds. It will be challenging, if not hopeless, to observe
in those decays the mixing effects related to the neutrino mass
generation.

\vspace{0.5cm}
\noindent %
{\bf Acknowledgement}

This work is supported in part by Grants No. NSFC-10775074, No.
NSFC-10975078, No. NSFC-11025525, and the 973 Program 2010CB833000.
YL would like to thank T. Plehn for the invitation for a visit and
Institut f\"ur Theoretische Physik, Universit\"at Heidelberg for
hospitality, where this work was done at its final stage. We thank
the anonymous referee who suggested including the constraint from
muon-electron conversion in nuclei, which turns out to be very
stringent.

\noindent %

\end{document}